\documentclass[twocolumn,pra,showpacs,superscriptaddress]{revtex4}
\usepackage{graphicx}
\begin{document}

\newcommand{\refeq}[1]{Eq.~(\ref{#1})}
\newcommand{\reffig}[1]{Fig.~\ref{#1}}
\newcommand{\refsec}[1]{Sec.~\ref{#1}}

\newcommand{\showfig}[1]{\includegraphics[width=0.47\textwidth]{#1}}

\title{Spinful bosons in an optical lattice}
\author{Sara Bergkvist}
\affiliation{Condensed Matter Theory, Physics Department, KTH, AlbaNova
  University Center, SE-106 91 Stockholm, Sweden}
\email{sarber@kth.se}
\author{Ian McCulloch}
\affiliation{
Institute for Theoretical Physics C, RWTH Aachen University, 
D-52056 Aachen, Germany}
\author{Anders Rosengren}
\affiliation{Condensed Matter Theory, Physics Department, KTH, AlbaNova
  University Center, SE-106 91 Stockholm, Sweden}


\date{\today}

\begin{abstract}
We analyze the behavior of cold spin-1 particles with
antiferromagnetic interactions in a one-dimensional optical lattice
using density matrix renormalization group calculations. Correlation
functions and the dimerization are shown and we also present results
for the energy gap between ground state and the spin excited states.
We confirm the anticipated phase diagram, with Mott-insulating regions
of alternating dimerized $S=1$ chains for odd particle density versus
on-site singlets for even density.  We find no evidence for any
additional ordered phases in the physically accessible region, however
for sufficiently large spin interaction, on-site singlet pairs
dominate leading, for odd density, to a breakdown of the Mott
insulator or, for even density, a real-space singlet superfluid.
%We see no evidence for any additional phases due to the spin interaction,
%in particular the even density superfluid state apparently contains
%no novel ordering such as a real-space paired superfluid.
%The expected phase diagram with a dimerized
%phase in the insulating regions with an odd particle density and
%on-site singlets in the insulating regions with even density is
%confirmed.

\end{abstract}

\maketitle

\section{Introduction} 
Strongly interacting systems which are at the forefront of study in
theoretical condensed matter physics can be realized with cold atomic
gases in optical lattices.  Spinor atoms in optical lattices
constitute a novel realization of quantum magnetic systems. These new
realizations have several advantages compared to their 
condensed matter counterparts. One is precise knowledge of the underlying
microscopic model, another is the possibility to control the parameters of
the lattice Hamiltonian, a third is the absence of impurities in the
system. Due to the smallness of the scattering length compared to
inter-particle separation, a gas of degenerate alkali-metal atoms is
considered as a weakly interacting gas\cite{Pethick}. However this
is no longer true if the atomic scattering length is changed by means
of a Feshbach resonance\cite{BuNa02} or if an optical lattice created by
standing wave laser beams is used to confine the atoms in the minima
of the lattice potential, which strongly enhances the effects of the
interactions.

For the case of the optical lattice the Mott insulator-superfluid
quantum phase transition was first demonstrated in the seminal
experiment of Greiner et al.\cite{GrNat02}. for $^{87}$Rb. This
experiment has led to an enormous research activity in the field of
cold atoms confined in optical lattices.  In most experiments on
cold atoms a magnetic field is used to create a trap that confines the
atoms in the lattice. This means that the spins of the atoms are 
fully polarized
so that the atoms behave as spinless particles.
Recently a new experimental setup that uses an optical, instead of
magnetic, trap has been developed\cite{Bloch1,Bloch2}. Using
this type of confinement atoms with different spin polarizations are
trapped and the scattering of the atoms become spin
dependent\cite{DePRL02,ImPRA03,ZhAnn03}. The spin
interaction may be either antiferromagnetic or ferromagnetic in its
structure depending on the scattering length which is material
specific. For sodium $^{23}$Na, the interaction is 
antiferromagnetic\cite{DePRL02}.

A model for spinful bosons in optical lattices has been studied before
by a number of
groups\cite{DePRL02,ImPRA03,ZhAnn03,SnPRB04,YPRL03,Kimura,ImPRL04,ZhPRB04},
and a phase diagram has been predicted. For integer number of atoms
per lattice site there are two regimes. One, where the kinetic energy
is dominating over the potential energy, has a superfluid groundstate
and the other, where the interaction energy is the most important,
has a Mott insulating groundstate. For non-integer particle density
the superfluid groundstate is always prevailing. For the
one-dimensional case the predicted phase diagram has a spin-dimerized
phase in the insulating regions with odd particle density and on-site
spin singlets in the insulating regions with even density. This phase
diagram was recently confirmed numerically\cite{Rizzi}. However, for
the study of the order parameter the authors' resorted to a mapping of
the boson model to a spin model\cite{ImPRA03}. By doing so the
dimension of the local Hilbert space is reduced from 20 or more
(depending on the maximum number of bosons per site) to three,
and accordingly the calculations are less time and memory
consuming. However, this mapping is only applicable to the insulating
regions with odd density, and is valid in the limit of very strong
atom-atom repulsion.

Our aim with this article is to study the spinful model directly. The
correlation functions and excitation energies are presented for the
first three insulating regions. We also compare results for the
insulating systems with those for the superfluid phase. All
calculations in this article are done using density matrix
renormalization group (DMRG) with open boundary conditions.

The outline of the paper is as follows; In \refsec{Model} the model
and the mapping from the spinful Hamiltonian to the spin-one chain is
presented. In \refsec{Method} some details about the calculations
are given. In \refsec{Corr} we present particle-particle and
spin-spin correlation functions for the system. In \refsec{Dim} the
dimerization in the spin-spin correlation is obtained and from this it
is concluded that the third Mott lobe is dimerized and that the first
lobe most probably is dimerized. In \refsec{Energy} the energy gap
to excited spin states is investigated. It is apparent that the second
Mott lobe has on-site singlets and that the gap decays with a
universal behavior when the superfluid phase is approached. We also
show that the odd density insulating systems have a small spin
gap and that this energy gap is non-zero for all spin-interaction
strengths. Section \ref{Singlet} contains a discussion on the
conditions under which bound on-site singlet phases occur.
This gives several new states, including a singlet insulator
and a superfluid phase
of condensed singlet pairs (SSC)\cite{DePRL02,ZhAnn03}.
In this latter phase the tunneling of individual atoms
is suppressed and only singlet pairs
tunnel\cite{DePRL02}. The last section, \refsec{conclusion}, is
devoted to a summary and conclusions.

\section{The model}
\label{Model}

\begin{figure}
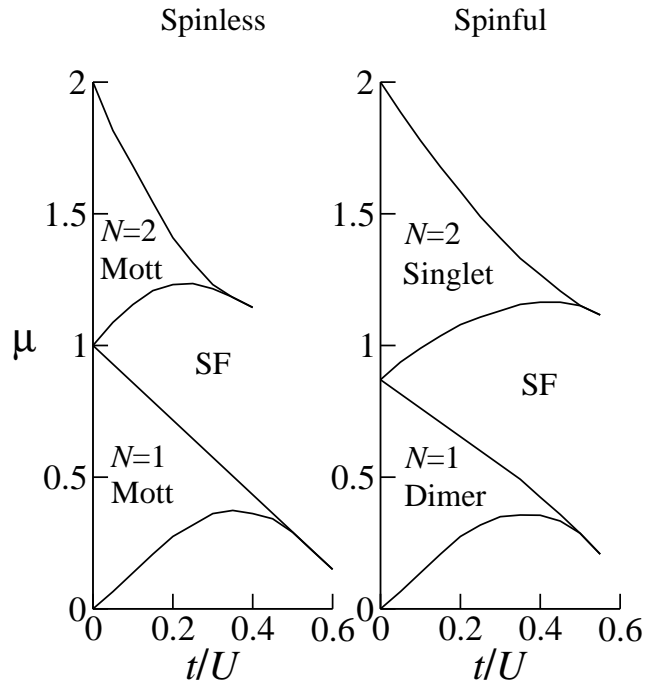

\centering
\showfig{Fig1.eps}
\caption{A sketch of the phase diagram of the Bose-Hubbard
model\protect{\cite{Rizzi}}. The left panel shows the Mott lobes for the
spinless Bose-Hubbard model and the right panel shows the spinful
phase diagram ($J/U=0.1$).}
\label{phaseD}
\end{figure}

In most experiments on optical lattices a magnetic trap is used to
confine the atoms. Due to the spin dependent interaction between the
atoms and the trap only one of the spin species is confined in the
lattice. This system is very well described by the Bose-Hubbard
model\cite{JaPRL98}. The phase diagram for the spinless Bose-Hubbard
model was discussed in the seminal work by Fisher et al.\cite{FiPRB89}
using a scaling theory and renormalization-group calculations. They
also derived the exact phase diagram within mean-field theory,
i.e.~for an infinite-range hopping model.  The phase diagram for the
one-dimensional model has been obtained numerically with a high
precision by K\"uhner et al.\cite{KuPRB00,KuPRB98} using
the DMRG method. The phase diagram has Mott insulating regions (Mott
lobes) and superfluid regions. A sketch of the phase diagram is shown
in the left panel of \reffig{phaseD}.

When an optical trap is used instead of a magnetic trap, atoms with
all spin polarization are trapped, with alkali atoms having hyperfine spin
$F=1$. The scattering between two atoms takes place in the spin-zero
and the spin-two sectors. Since the scattering length is different
in these two sectors the on-site repulsion becomes spin dependent,
resulting in the following Hamiltonian\cite{DePRL02,ImPRA03},
\begin{equation}
\begin{array}{rcl}
H&=&\displaystyle -t\sum_{i,\sigma} 
\left( b^+_{i,\sigma}b^{}_{i+1,\sigma}+ b^{}_{i,\sigma}b^+_{i+1,\sigma} \right)\\
& & \quad
+\displaystyle \sum_i \left( U \; n_i(n_i-1)+J \; {\bf S}_i^2 \right) \;.
\end{array}
\label{EqHam}
\end{equation}
Here $b^+_{i,\sigma}$ and $b_{i,\sigma}$ are the creation and
annihilation operators for bosons with $z$-component of spin $\sigma=0,\pm 1$,
in the lowest Bloch band localized on site $i$,
$n_i=\sum_{\sigma}b^+_{i,\sigma}b_{i,\sigma}$ is the local density,
$t$ is the hopping integral for atomic wavefunctions between different lattice
sites, $U$ is the usual on-site Coulomb repulsion,
$\mathbf{S}_i=\sum_{\sigma,\sigma'}b^+_{i,\sigma}\mathbf{T}_{\sigma,\sigma^{'}}
b_{i+1,\sigma^{'}}$ is the spin operator for spin one particles
at lattice site $i$, with
$\mathbf{T}_{\sigma,\sigma^{'}}$ being the usual spin-1 matrices.
$J$ is the spin dependent
interaction whose value and sign depend on the material. For $^{23}$Na,
$J$ is positive, i.e., the spin interaction is antiferromagnetic, and
$J/U\approx 0.04$\cite{ImPRA03}. In this paper we focus mainly on
the parameter range appropriate for $^{23}$Na.

The insulating regions with an odd density form an effective
spin-one chain, as the spins on each lattice site will combine to the
lowest possible spin. This maps the insulating odd density
region for the spinful Bose-Hubbard model onto the bilinear-biquadratic
spin-one chain with
Hamiltonian\cite{ImPRA03,YPRL03,ZhAnn03,Rizzi}
\begin{equation}
H=\sum_{\langle ij\rangle}(\cos\theta({\bf S}_i \cdot{\bf S}_j)+
\sin\theta({\bf S}_i \cdot{\bf S}_j )^2)
\label{Eq2}
\end{equation}
where $\tan(\theta)=\frac{1}{1-2J/U}$, $\theta\in[-3\pi/4,-\pi/2]$,
for the first Mott lobe. Similar expressions should exist for higher (odd)
Mott lobes.

The mapping is obtained by making a lowest order, i.e.~second
order, perturbation expansion in $t$ of the
Hamiltonian~\refeq{EqHam}. To second order only pairwise
interactions between atoms on neighboring sites are generated. For the
first Mott lobe, i.e., in the insulating state with exactly one boson
per site, the nearest-neighbor interactions are always of the form
given by \refeq{Eq2}. However, the derivation of the $\theta$
dependence on $J$ and $U$ is only valid in the limit $t \ll U$. 
Away from the limit $t \ll U$ but still in the insulating phase, higher 
order terms
have to be included which will, in addition to renormalizing the
nearest-neighbor couplings, add spin couplings beyond nearest-neighbor.
 
Despite many years of study, the phase diagram of the spin 1 chain
is not fully resolved.
%The phase diagram for the spin chain is debated. 
For $\theta=0$, the
antiferromagnetic Heisenberg model, the spectrum has a finite
Haldane gap\cite{HaPRL83}, whereas at $\theta=\pm\pi$ the
system is ferromagnetic. Actually, both phases have a finite extension
in parameter space. Ferromagnetism exists for $-\pi <\theta<
-3\pi/4$ and $\pi/2<\theta<\pi$ while the massive Haldane phase
exists for $-\pi/4<\theta<\pi/4$. At the lower end of this interval,
$-\pi/4$, the gap vanishes, but opens again for $\theta< -\pi/4$ and a
massive dimerized state with spontaneously broken translation symmetry is
found. For the special point $\theta=-\pi/2$, the ground-state energy,
gap, correlation length and the dimer order parameter can be calculated
exactly\cite{DimerOrder}.
The unresolved issue is whether the dimerized phase
extends all the way to $\theta=-3\pi/4$ or if there exists
another phase in between the dimerized and the ferromagnetic phase near
$\theta=-3\pi/4$.

Chubukov\cite{ChubDimer}, studying fluctuation effects near the end
point of the ferromagnetic phase, concluded that the dimerized and the
ferromagnetic phase are separated by a disordered phase, a gapped
non-dimerized nematic.  According to Chubukov a direct transition
between the dimerized and ferromagnetic phases is very unlikely since
completely different symmetries are broken in these phases. Chubukov
claimed that the dimer order parameter and the gap vanish
simultaneously at a $\theta_c$ above, but close to,
$\theta=-3\pi/4$. For $\theta<\theta_c$ the gap opens up and closes
again at $\theta=-3\pi/4$ whereas the dimer order parameter is zero in
this interval. Buchta et al.~recently performed highly accurate DMRG
calculations\cite{BuPRB05} for this region.  Their results indicate
that the dimer phase prevails down to $\theta=-3\pi/4$, although they
cannot rule out that a non-dimerized phase exists in an extremely
small interval of $\theta$ close to $\theta=-3\pi/4$. 
A recent preprint by  L\"auchli, Schmid and Trebst\cite{Lauchli},
use a strong-coupling series expansion which gives a vanishing
gap for $-3\pi/4<\theta<-0.67\pi$, but they also present DMRG results
that are consistent with those of Buchta et al.\cite{BuPRB05}
showing a small but finite gap in this region.
%Another
%calculation, by L\"auchli, Schmid and Trebst\cite{Lauchli}, combined
%DMRG with exact diagonalization and found evidence for a critical
%phase, i.e.~gapless and non-dimerized, with quadrupolar correlations
%in the interval $-3\pi/4<\theta<-0.67\pi$.  This interval is however
%much too wide to be compatible with the recent DMRG calculations of
%Buchta et al.\cite{BuPRB05}. 
Porras et al.\cite{Verstraete} introduced
a new numerical algorithm and applied it to the same problem. They
found indications of a quantum phase characterized by nematic
quasi-long-range order in the interval
$-3\pi/4<\theta<-0.7\pi$. However the accuracy of this calculation is
difficult to quantify; their results are also consistent with either
the correlation length being longer than the size of the chains
considered, or a gap that is smaller than the numerical
accuracy\cite{Verstraete}. Also, Rizzi et al.\cite{Rizzi} concluded
from DMRG calculations that there is no intermediate nematic phase,
although they found indications that a tendency towards nematic
ordering is enhanced as the dimer order parameter goes to zero.

In the odd density insulating state with more than one boson 
per site there is an
additional constraint in order to arrive at the
Hamiltonian \refeq{EqHam}: configurations with spin on individual
sites higher than one have to be neglected. Matrix elements for
scattering into such states are of the order $(nt)^2/U$ and their
energy is set by $J$, where $n$ is the number of bosons per
site. Therefore in order to obtain the bilinear-biquadratic Hamiltonian, 
the condition is that $nt \ll (UJ)^{1/2}$\cite{ImPRA03}.

The insulating regions with an even density behave completely
differently compared to those with odd density. For an even number
of spins per site, the particles form bound states of singlet pairs 
resulting in a rather large energy gap to the excited states. 
Note however that unless the spin interaction
is very large, the even density superfluid shows no such real-space pairing.

\begin{figure}
\centering
\showfig{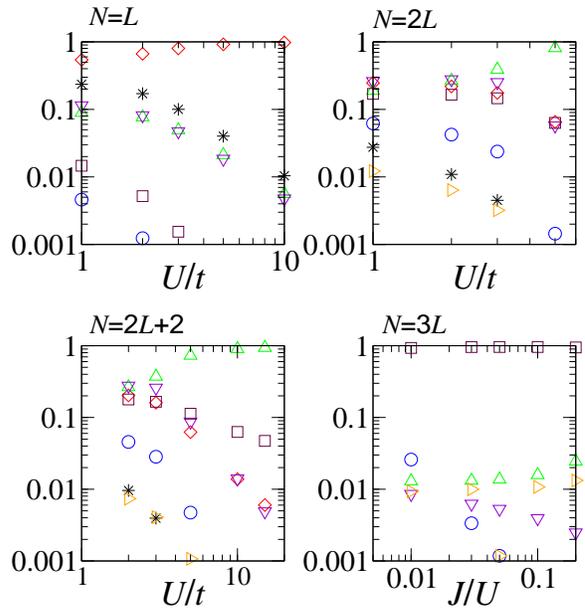}
\caption{(Color online) The population of the single-site basis states for an average
density of one, two, $N=2L+2$ and three particles. The probability for
the different sectors is presented as a function of the on-site
repulsion $U/t$, for $J/U=0.05$, in the first three panels. The last
panel shows the third Mott lobe as a function of $J/U$, with
$U/t=15$. The different sectors, $(n,S)$, presented in the figure are:
star(0,0), diamond(1,1), triangle up(2,0), triangle down(2,2),
square(3,1), circle(3,3), triangle right(4,0).}
\label{besatt}
\end{figure}

The spin interaction affects the critical interaction strength for the
transition from the Mott insulating to the superfluid phase. In the
Mott lobes with even particle densities, i.e.~the insulating regions
with on-site spin singlets, the insulating regions are stabilized by
the spin interaction. On the other hand the Mott lobes with an odd
density are weakened by the spin interaction, the density fluctuation
is enhanced since the particle hopping is the process that carries the
spin-interaction\cite{TsPRA04}. In the right panel of
\reffig{phaseD} a sketch of the new phase diagram is shown together
with the phase diagram for the spinless model (left panel).

\section{Method}
\label{Method}

The method we have used is the Density Matrix Renormalization 
Group\cite{WhiteDMRG,UliReview05}, adapted to exactly handle the
non-Abelian $SU(2)$ symmetry\cite{NonAbelianDMRG02}. The starting point
of this formulation is the Wigner-Eckart theorem\cite{SU2Intro},
which states that the matrix elements of $SU(2)$-invariant operators
factorize into a component that is \textit{independent} of the
geometry (i.e., the quantization axis), and a geometrical factor
(Clebsch-Gordan coefficient). In particular, we have the relationship
\begin{equation}
\langle j' m' | A^{[k]}_M | j m \rangle
= C^{j}_{m} {}^{k}_{M} {}^{j'}_{m'} \; \times \; 
\langle j' || \mathbf{A}^{[k]} || j \rangle \; ,
\label{eq:wignereckart}
\end{equation}
where  $\langle j' || \mathbf{A}^{[k]} || j \rangle$ is the
\textit{reduced matrix element} of the rank $k$ tensor $\mathbf{A}^{[k]}$.
From this relation, all of the operations required for the DMRG
(tensor products, wavefunction transformations etc.), can be written
in terms of the reduced matrix elements only, with an appropriate
$6j$ or $9j$ coupling coefficient\cite{NonAbelianDMRG02,SU2Intro}.

The use of symmetries to improve the efficiency of DMRG calculations
was recognized from the beginning\cite{WhiteDMRG}, and presumably arose
from the Wilson's use of spin in NRG\cite{WilsonNRG},
the historical forerunner of DMRG. The utility of \textit{Abelian}
symmetries (typically $U(1)$, such as particle number, or $z$-component
of spin), is that it imposes a sparseness constraint on the operator
matrix elements; given a basis labeled by the particle number, for example,
the matrix elements of
some irreducible operator $\langle n' | A^{[N]} | n \rangle$ are non-zero
only for $n' = n+N$.
By storing only the matrix elements that are
permitted by symmetry to be non-zero, the memory use and computational
cost is greatly improved. Note however that \textit{a priori} this
has no effect on the accuracy, and in principle for a given basis size
the obtained matrix elements and wavefunction will be identical whether
the symmetry constraint is used or not. This is in contrast
to the case of non-Abelian symmetries, where \refeq{eq:wignereckart}
implies a \textit{reduction} in the basis size needed to represent
the Hamiltonian matrix elements. That is, a single basis state of total spin
$j$ requires $2j+1$ states to represent in a $U(1)$ basis. This results
in a proportional reduction in the basis size $m$ required for a given
accuracy.  Depending on the spin of the block basis states
this reduction factor typically ranges from around 3 in the 
Mott insulating phase
where the block spins tend to be minimal, to 8 or so in the superfluid
phase. Since the computational cost
of the algorithm is proportional to $m^3$, this results in orders
of magnitude improvement in the efficiency.

For these calculations, we typically used 350 basis states (equivalent
to approximately $1000-2500$ basis states of a $U(1)$ calculation).
For the finite size scaling analyses we used up to 300 lattice 
sites for the phases with total number of bosons $N$ equal
to the number of lattice sites $L$, up to 120 sites for $N=3L$ and
up to 70 sites for the $N=2L$ superfluid.
We fixed the maximum number of bosons per site to be 5, except
for the $N=L$ superfluid where we found four bosons per site to be
sufficient, and the $N=L$ insulator where we found 3 bosons per site 
to be sufficient.
This is justified by the occupation number shown in
\reffig{besatt} where the average number of bosons in the different
single site basis states are shown.

\section{Correlation functions}
\label{Corr}

\begin{figure}
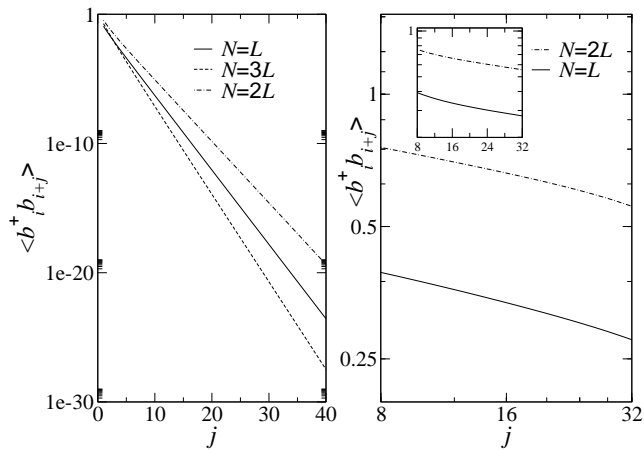

\centering
\showfig{Fig3.eps}
\caption{The $\langle b^+b\rangle$-correlation
function. The left hand panel is in the insulating regime
($U/t=10$, $U/t=10$, $U/t=15$ for $N=L$, $N=2L$, $N=3L$ respectively),
the right hand panel shows the superfluid ($U/t=1$ for both $N=L$ and
$N=2L$), log-log plot (Inset: same data with a log-linear scale).
We fix $J/U=0.05$ in all cases. }
\label{FigBB}
\end{figure}

In this section we discuss the calculated hopping correlation
functions and the spin-spin correlation functions. In the superfluid
the hopping correlation function $\Gamma(j)=\langle
b^+_{i,\sigma}b_{i+j,\sigma}\rangle$ decays with a power-law as 
we expect\cite{FiPRB89}. The low-energy
energy physics of the superfluid phase is described by a Luttinger
liquid and the decay of the hopping correlation function
$\Gamma(j)\sim j^{-K/2}$ where $K$ is the Luttinger liquid
parameter\cite{HaPRL81}. Since we use open boundary conditions,
measures have to be taken to reduce the effects of the boundary. The
most important effect is oscillations in the local density. In the
superfluid phase they show the characteristic Luttinger liquid
power-law decay away from the edge of the system. These density
fluctuations will affect the hopping correlation function
$\Gamma(j)$. Since we are interested in the properties of the infinite
system we reduce the effects of these fluctuations by averaging over
pairs of $\langle b^+_{i,\sigma}b_{i+j,\sigma}\rangle$ for fixed $j$.  
We discard
contribution from the ten lattice sites closest to the boundary where
the Friedel oscillations are largest. Beyond this distance the
numerical behavior of $\Gamma$ is no longer sensitive to the boundary
oscillations.  The right panel of \reffig{FigBB} shows results for the
hopping correlation function for two different integer densities
displaying superfluid behavior.

In the Mott insulating phase the hopping correlation function
$\Gamma(j)$ is decaying exponentially\cite{FiPRB89}, where the decay
rate is determined by how close we are to the phase boundary of the
Mott phase. As one would expect, the boundary effects decay
exponentially in this region. 
The left panel of \reffig{FigBB} shows results for the
hopping correlation function for three different integer densities.

\begin{figure}
\centering
\showfig{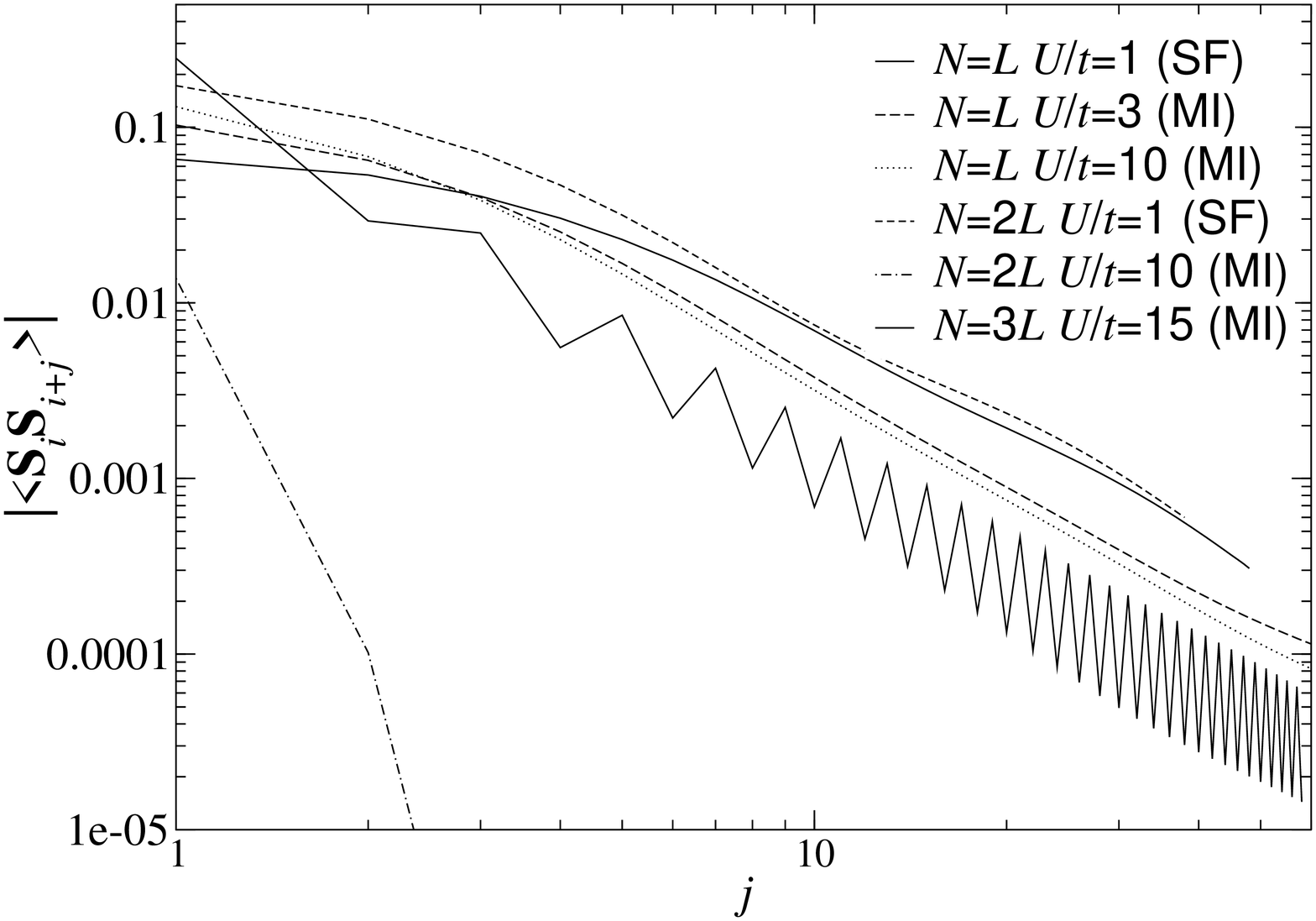}
\caption{The spin-spin correlation for $N=L$, both insulating and
superfluid, $N=2L$ both insulating and superfluid and $N=3L$
insulating, $J/U=0.05$. The clear oscillation in the correlation
function for the third Mott lobe arises from the large dimerization.}
\label{FigSS}
\end{figure}

In \reffig{FigSS} the spin-spin correlation functions for different
systems are presented for Mott insulating and superfluid regions. The
spin interaction strength is set to $J/U=0.05$. The figure demonstrates
that the correlation function is rather independent of the phase for
the odd density systems. It decays in the same way for all the odd
density systems and also for the superfluid system with an even
density. The insulating system with a density of three atoms per site
is dimerized, for this interaction strength, which is the reason for the
oscillating structure of the spin-spin correlation.

For the insulating system with an even density the spin-spin
correlation is exponentially decaying, consistent with the picture
of on-site singlets.

\section{Dimerization}
\label{Dim}

\begin{figure}
\centering
\showfig{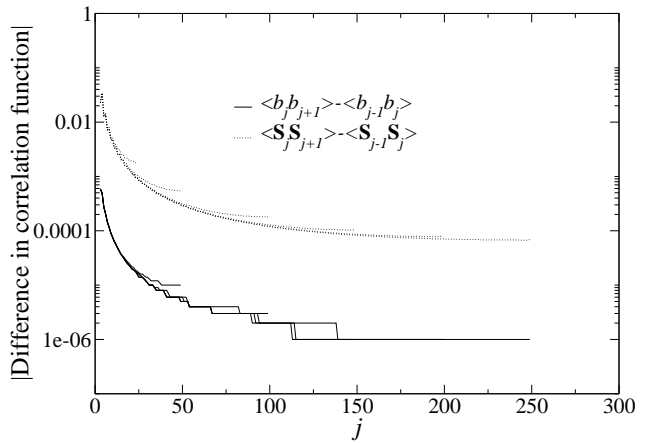}
\caption{The absolute value of the difference in neighboring
correlation functions in the first Mott lobe ($U/t=10$, $J/t=0.5$) for
different system sizes. The dimerization is calculated up to the
middle of the chain.}
\label{FigDim10_05}
\end{figure}

One of the questions that we try to resolve in this article is the
nature of the spin order in the first Mott lobe. 
The phase is, as mentioned above in
\refsec{Model}, believed to be dimerized.  The standard way to
characterize the dimerization is by using a dimer order parameter
which usually is defined as the difference in bond energy in the
middle of the chain
\begin{equation}
D=\langle H_{j,j+1}\rangle-\langle H_{j-1,j}\rangle.
\end{equation}
In the spinful Bose-Hubbard model this dimer order parameter is given
by $\langle b_j^+b_{j+1}-b_{j-1}^+b_j\rangle$ since all other terms in the
Hamiltonian are on-site. This difference is very difficult to measure
since it is very small, see \reffig{FigDim10_05}. The figure shows
the absolute value of the mentioned difference  as a function
of the lattice site, $j$, for a Mott insulating system. In the
figure $j$ runs between 1 and the middle of the chain. The
dimerization is calculated for a couple of different system sizes to
illustrate the boundary effects. The bond energy is indeed oscillating
between different lattice sites but the values are so small that it is
difficult to tell whether or not the oscillation will be non-zero in
the thermodynamic limit. The open boundary conditions break the
translational symmetry which is necessary to see a dimerized order
parameter. For periodic boundary conditions the groundstate would be a linear
combination of the two possible degenerate dimerized configurations and
translational symmetry would be restored for any finite $L$.

The dimerization in the spin-spin correlation functions for the Mott
insulating system has also been studied, the result is presented in
\reffig{FigDim10_05}. In this figure the dimerization in the
spin-spin correlation is shown as a function of the lattice site, for
a few system sizes. 

\begin{figure}
\centering
\showfig{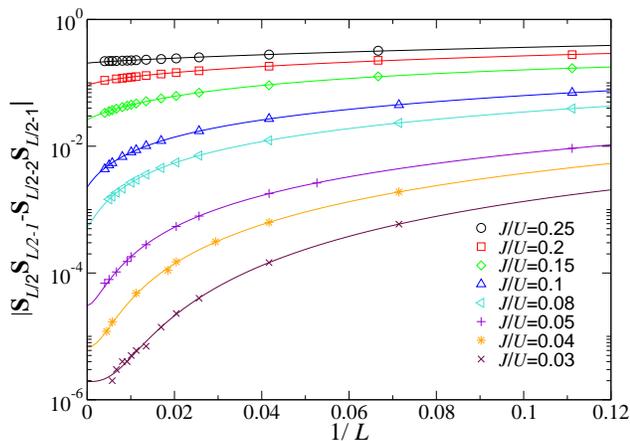}
\caption{(Color online) The dimerization in the spin-spin correlation function as a
function of the inverse lattice size for the first Mott lobe ($U/t=10$),
for different strength of the spin interaction. The solid lines
are the finite size scaling.}
\label{FigDimMott1}
\end{figure}

\begin{figure}
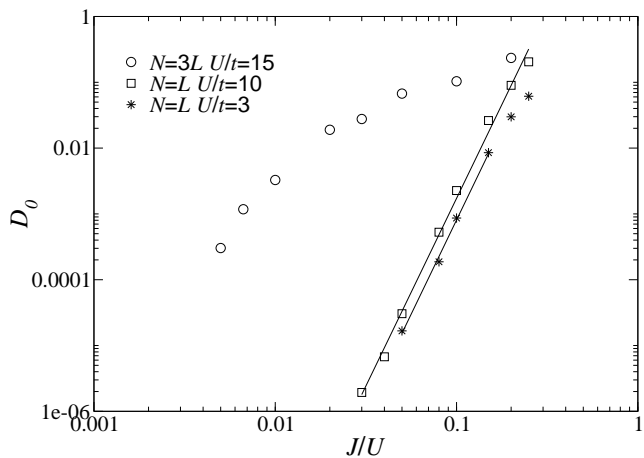

\centering
\showfig{Fig7.eps}
\caption{The dimerization in the thermodynamic limit as a function of
spin interaction strength. The dimerization in the first Mott lobe
decays algebraically with a fitted exponent $\sim 5.7$ and there is no
suggestion that it will be non zero before $J/U=0$. In the third Mott
lobe the dimerization is much larger.}
\label{FigD0}
\end{figure}

(Color online) The dimerization in the spin-spin interaction for different
interaction strengths in the first Mott lobe is shown in
\reffig{FigDimMott1}. In the figure the spin-spin dimerization at
the middle of the lattice is shown as a function of the system
size. There is a dimerization when the spin interaction becomes large
but it is hard to draw any conclusions about the thermodynamic limit
for small interaction strengths. We have done a finite size scaling
of the points with the following fitting function\cite{BuPRB05}
\begin{equation}
D=D_0+dN^{-\beta}\exp(-N/2\xi)
\end{equation}
The infinite size values for the first Mott lobe are presented in
\reffig{FigD0}. The dimerization goes to zero algebraically with an
exponent $\sim 6$, i.e., $D_0\propto (J/U)^6$. 
This is in agreement with a calculation on the
bilinear biquadratic chain\cite{Rizzi}. The curve shows no indication
of a non-zero dimerization before $J=0$.  The values in the figure are
obtained for $U/t=10$ and for $U/t=3$. The fact that the curves for
$U/t=3$ and $U/t=10$ fall on top of each other shows that the
dimerization is present in the entire Mott lobe.  This conclusion
cannot be deduced from the bilinear-biquadratic chain, since the
mapping is only valid for small hopping.

In \reffig{FigD0} the dimerization in the spin-spin correlation for
the third Mott lobe as a function of spin interaction strength is also
shown.  The values are infinite size values obtained in the same
manner as for the first Mott lobe. However, the system sizes are
smaller so the error in finite size scaling is larger.  As was
mentioned in \refsec{Model} also for the third Mott lobe the spinful
Bose-Hubbard model could be approximately mapped to the
bilinear-biquadratic spin chain. However in this case the constraints
for the perturbation expansion in $t$ to be valid are more severe,
configurations with spin on individual sites higher than one have to
be neglected leading to the condition $nt \ll (UJ)^{1/2}$. For the
dimerization in the third Mott lobe presented in \reffig{FigD0} this
condition reads $J/U\gg 0.04$. As configuations with spin higher than
one come into play we see a possible cross over in the behavior and a
down-turn in the curve for $D_0$ for $J \leq 0.01$. However still we
find no indications of $D_0$ being zero for $J$ larger than zero.  The
third Mott lobe is definitely dimerized for $J/U=0.005$ and we
conclude that there is no extra phase between the dimerized phase and
the ferromagnetic in the third Mott lobe.

\section{Spin gaps}
\label{Energy}

\begin{figure}
\centering
\showfig{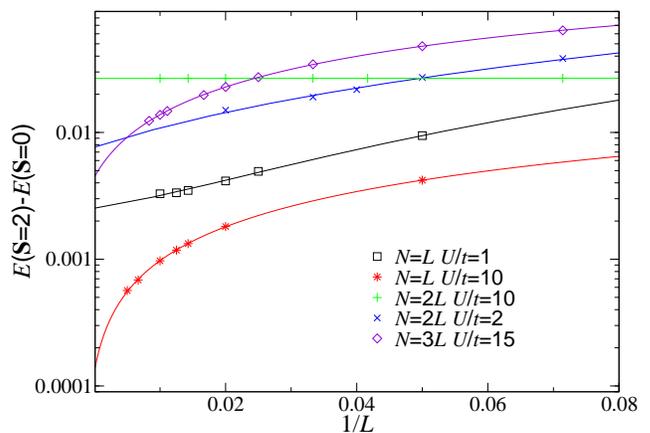}
\caption{(Color online) The difference in energy between states with $S=2$ and the
ground state, $S=0$, as a function of the inverse system size for
$U/t=10$ and different spin interaction strengths. The values for the
second Mott lobe (denoted `$+$') are scaled by 1/100.}
\label{FigEN}
\end{figure}

In this section, 
the difference in energy between the ground state, which is a spin
singlet, and the first excited spin state, which has spin $S=2$, is
studied. This difference is calculated for several system sizes and is
shown as a function of the inverse system size in \reffig{FigEN}. 
The energy gap to excited states in the first Mott lobe has been
calculated before with DMRG using the mapping to the
bilinear-biquadratic spin chain\cite{Rizzi} and also in the limit of
zero hopping\cite{ZhEPL03}.

\begin{figure}
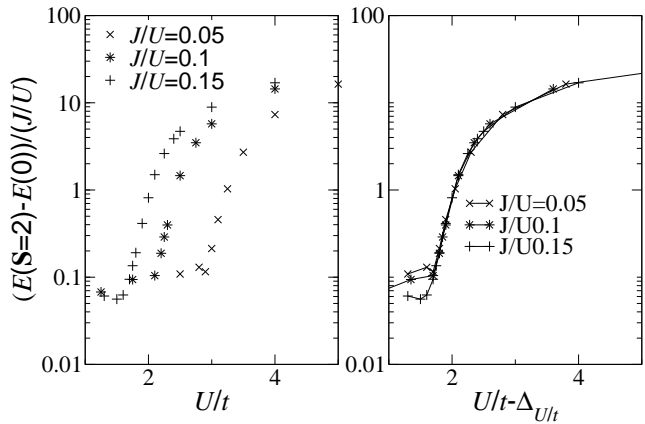

\centering
\showfig{Fig9.eps}
\caption{The spin gap in the thermodynamic limit for the second Mott
lobe. In the left panel the values are shown as a function $U/t$. In
the right panel the lines are shifted by a constant $\Delta_{U/t}$
along the $x$-axis
to show the similarity in the decay of the gap for the different
interaction strengths.}
\label{FigEMott2}
\end{figure}

%\begin{figure}
%\centering
%\showfig{Fig10.eps}
%\caption{ The energy density as a function of the magnetization $M =
%S/L$ for the second Mott lobe, there is a linear increase in energy
%as a function of the spin. }
%\label{FigESM2}
%\end{figure}

From the figure it is obvious that the decay of the energy gap, with
respect to the inverse system size, for the Mott insulating state with
two particles per site behaves differently to that of the other
cases. The gap is large which is consistent with the 
on-site-singlet picture and approaches the thermodynamic limit as
$L^{-2}$\cite{ZhEPL03}.

We have studied how the energy gap in the thermodynamic limit for the
second Mott lobe approaches the superfluid value as the on-site
repulsion, $U$, is decreased. The result is shown in
\reffig{FigEMott2}, for $J/U=0.05$, $J/U=0.1$ and $J/U=0.15$. To
obtain the values in the figure calculations are done for different
interaction strengths and system sizes. The infinite size values are
found by fitting second-order polynomials to the data points. In the
figure the infinite size values are presented as a function of the
on-site repulsion for three different strengths of the spin-spin
interaction. In the right panel of \reffig{FigEMott2} the abscissa
($U/t$) of two of the curves in the left panel has been shifted by a
constant ($\Delta_{U/t}$), different for each of them, to show that
the curves can be collapsed on top of each other. From this it is
clear that the energy gap decays in the same manner for the three
interaction strengths when the superfluid phase is approached.

\begin{figure}
\centering
\showfig{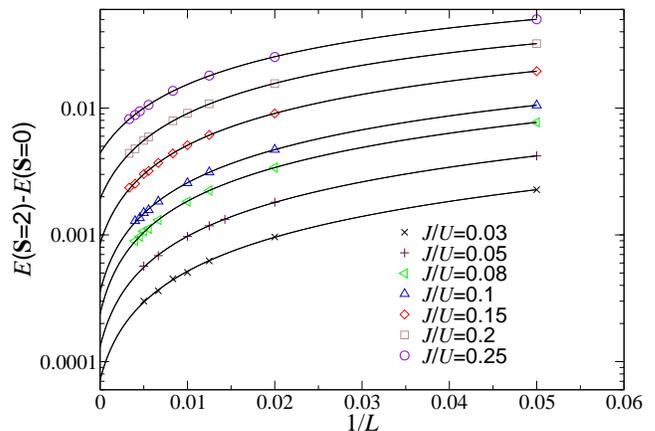}
\caption{(Color online) The difference in energy between states with $S=2$
and the ground state, $S=0$, as a function of the inverse
system size for $U/t=10$ and different spin interaction strengths the
solid lines are finite size scalings.}
\label{FigEL}
\end{figure}

\begin{figure}
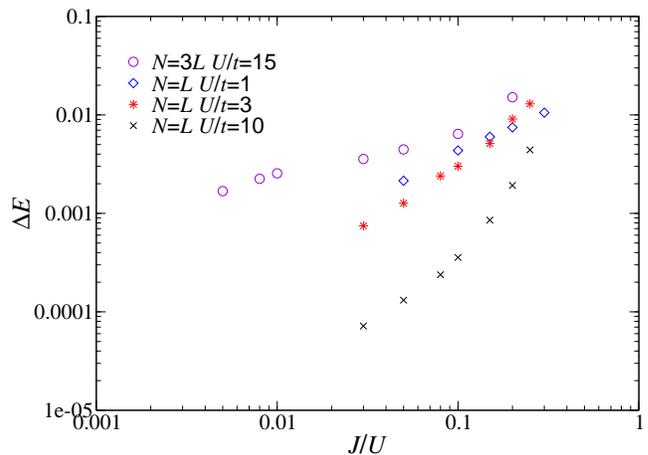

\centering
\showfig{Fig12.eps}
\caption{(Color online) The energy gap between the ground state and the first spin
excited state in the thermodynamic limit, as a function of the spin
interaction for odd density systems. The system with $U/t=1$ is superfluid.}
\label{FigEDiffU}
\end{figure}

We have further studied how the gap approaches zero for the odd
density insulating regions when $J/U$ approaches zero. 
In \reffig{FigEL}, we show for the first Mott lobe at $U/t = 10$, 
the gap as a function
of system size, for a variety of different spin couplings. The symbols
mark the numerical data and solid lines show the extrapolation
to the thermodynamic limit using a second order polynomial fit in $1/L$.

In the first Mott lobe the gap in the thermodynamic limit is very
small and decays to zero as the interaction approaches zero, see
\reffig{FigEDiffU}. In the figure, results for $U/t=10$, i.e.~far
inside the Mott lobe, are shown together with values for $U/t=3$ which
is fairly close to the phase transition. Results for the third Mott
lobe and the superfluid phase are also shown in the figure. There is
no indication of that the energy gap becomes zero before $J=0$. The
four curves in \reffig{FigEL} can best be fitted with a power law
decay and nothing suggests an intermediate phase in between the
dimerized phase and the $J < 0$ ferromagnetic phase.

\begin{figure}
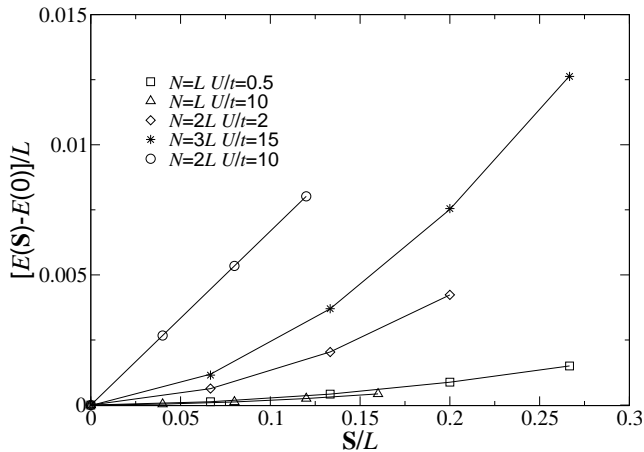

\centering
\showfig{Fig13.eps}
\caption{ The energy density as a function of the magnetization $M =
S/L$. The open circles for the second Mott lobe have been scaled
by a factor 20 to fit on the figure.  The fit for these points is
a straight line. All other fits are of the form
$E_0+kS(S+1)$, $J/U=0.05$. }
\label{FigES}
\end{figure}

In \reffig{FigES} we show the energy per site as a function of the
magnetization.  We show examples for both insulating and superfluid
systems. The spin excitation spectra for the first and the third Mott
lobe and the superfluid regions follow the parabolic form typical of
Heisenberg spin systems\cite{SchmidtParabolic},
\begin{equation}
E(S)=E_0+kS(S+1).
\end{equation}

The behavior of the spin excitations for the second Mott lobe is
rather different, see \reffig{FigES}. In this case, the energy
increases linearly with $S$ and with a steep slope, a result which is
consistent with the picture of the system being composed of on-site
singlets. The energy to break another singlet is independent of the
number of already broken singlets.
%The result for the second Mott lobe is not shown in the figure. For
%these systems the energy increases linearly with $S$ and with a steep
%slope.  A result which is consistent with the picture of the systems
%being composed of on-site singlets. The energy to break another
%singlet is independent of the number of already broken singlets.

The energy gap to excited spin states measures the magnetization
energy and as expected from the on-site singlet picture the second
Mott lobe has a higher magnetization energy.
% However due to spin
%conservation the system cannot be easily magnetized. The response of
%the three-dimensional system to a magnetic field has been studied
%before\cite{Kimura,ZhPRB04,ZhPRB05,SvPRA03}.
%It has been proposed and experimentally shown that this different
%response to an applied magnetic field may be used to find the
%locations of the Mott plateaus\cite{ImPRL04,Bloch2}.

\section{Singlet Condensates}
\label{Singlet}

In this section we comment on some of the proposed phases for the
spinful Hubbard model. Zhou and Snoek have suggested that a phase
exists in between the superfluid and the Mott insulating phases for
even particle density\cite{ZhAnn03}. In this hypothesized phase, the
particles would form a condensate of tightly bound on-site singlet
pairs. However, we see no evidence for this phase for any physically
realistic magnitude of spin interaction.  In \reffig{besatt}, we show
the occupation of each site basis state, for $J/U=0.05$. For this
value of the spin interaction, it is clear that there are no values of
the Coulomb repulsion $U$ where only even number of bosons is
favored. This also follows from a simple energy argument, that
suggests such a phase is unlikely: Since the spin interaction energy
is smaller than the charging energy it is costly to increase the
density on a site by two. The energy gain by keeping the spins coupled
to a singlet is much smaller than the energy penalty for the on-site
Coulomb repulsion.
 
\begin{figure}
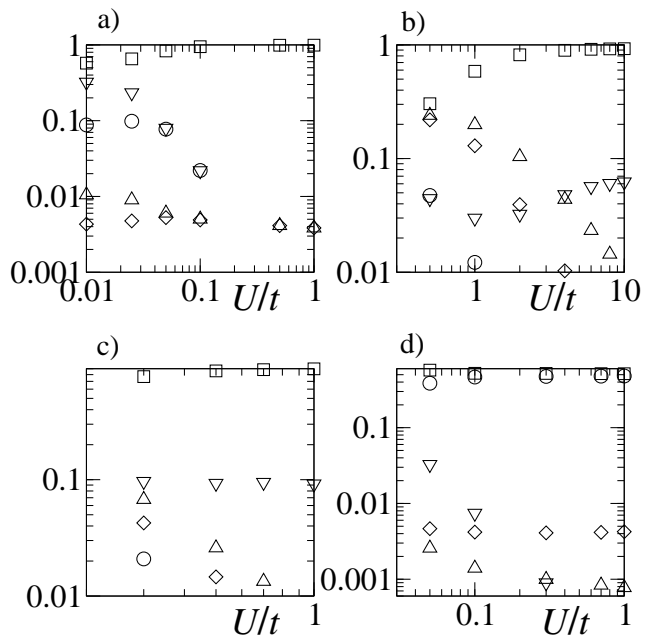

\centering
\showfig{Fig14.eps}
\caption{The probability to have a certain number of particles and
spin on a lattice site is presented for different systems and
different density and spin (n,S)sectors, circle(0,0), diamond(1,1),
square(2,0), triangle up(3,1), triangle down(4,0). a)$N=2L$ and
$J/t=10$: once $U$ and $t$ are comparable in size a singlet condensate
forms. b)$N=2L$ and $J/U=1$: for large $U$ this is an insulating
spin-singlet state, seen from the comparatively high probability in
the $n=4$ sector. c)$N=2L+2$ and $J/U=10$: spin singlet
condensate. d)$N=L$ and $J/t=10$: in this spin singlet condensate the
lattice is half-filled with spin singlets.  }
\label{besatt2}
\end{figure}

For large coupling the system behaves similarly to a
spinless Bose-Hubbard model populated with spin singlets. This limit
is tested and indeed for $J/t=10$ and $N=L$ the system is essentially
composed of $L/2$ empty sites and $L/2$ double occupied sites,
see \reffig{besatt2}.
The limit $J \gg U,t$
could localize the singlets to give a crystal phase\cite{DePRL02},
but we have not investigated this scenario.
In this figure we also show, as a function of $U$, the
population in the different on-site number and spin sectors for $N=2L$
particles per site for the spin interaction strength $J/t=10$, 
and also the superfluid system with $N=2L+2$ for $J/U=10$. 
As $J/U$
is reduced, the charging energy becomes equal to the spin energy and
the singlets are no longer bound.

\section{Conclusion}
\label{conclusion}

We have studied the 1D spinful Bose-Hubbard model using DMRG. By
examining the energy and correlation functions we have resolved the
behavior of the model for all of the parameter ranges that are
physically relevant for optical trapping experiments.

Much evidence is presented showing that the insulating system with an
even density is well described by the on-site spin-singlet
picture. The spin gap in the thermodynamic limit is rather
large, paired with a short correlation length for the
spin-spin correlation function.
The energy increases linearly with the spin of the system,
showing that the energy required to break another singlet is
given by the energy gap and is independent of the number of already
broken singlets, thus the singlets are strongly localized.

It has been more difficult to determine the nature of the other
insulating regions. From the particle-particle correlation function it
is concluded that there is a superfluid and an insulating phase just
as for the spinless case. It has previously been suggested that
these insulating regions are dimerized. The dimer order parameter
which uses differences in bond energy is difficult to measure in
numerical simulations since it is rather small. Instead the
oscillations in the nearest neighbor spin-spin correlation were used
to show that it is likely that the odd density per site Mott lobes are
always dimerized. This is a relatively strong result for the third
Mott lobe, where the dimerization order parameter is quite large.  For
the first Mott lobe our results are consistent with the $S=1$
Heisenberg calculations of Buchta et~al.\cite{BuPRB05} and Rizzi et
al.\cite{Rizzi}. In general the numerical accuracy of calculations in
this regime will be much better for the Heisenberg model compared with
the spinful Bose-Hubbard model, so our result does not contribute
substantially to the argument over the absence (or otherwise) of the
Chubukov nematic phase. However what is clear is that the dimerization
in the first Mott lobe shows scaling behavior with an exponent $\sim
6$ throughout the entire Mott lobe. 
We found no evidence for novel superfluid phases; in particular the on-site
paired superfluid suggested by Zhou and Snoek\cite{ZhAnn03} is absent in
this model.

The energy gaps to spin excited states were studied, showing that
the odd density (superfluid and insulating) and even density
superfluid phase follow the form of a parabolic
spectrum, proportional to $S(S+1)$.

\begin{acknowledgments}
S.B. is grateful to U. Schollw\"ock and W. Hofstetter for fruitful
 discussions and hospitality. This work was supported by the G\"oran
 Gustafsson foundation and the Swedish Research Council.
\end{acknowledgments}

% BibTeX-generated bibliography, some entries modified by hand


\begin{thebibliography}{37}
\expandafter\ifx\csname natexlab\endcsname\relax\def\natexlab#1{#1}\fi
\expandafter\ifx\csname bibnamefont\endcsname\relax
  \def\bibnamefont#1{#1}\fi
\expandafter\ifx\csname bibfnamefont\endcsname\relax
  \def\bibfnamefont#1{#1}\fi
\expandafter\ifx\csname citenamefont\endcsname\relax
  \def\citenamefont#1{#1}\fi
\expandafter\ifx\csname url\endcsname\relax
  \def\url#1{\texttt{#1}}\fi
\expandafter\ifx\csname urlprefix\endcsname\relax\def\urlprefix{URL }\fi
\providecommand{\bibinfo}[2]{#2}
\providecommand{\eprint}[2][]{\url{#2}}

\bibitem[{\citenamefont{Pethick and Smith}()}]{Pethick}
\bibinfo{author}{\bibfnamefont{C.}~\bibnamefont{Pethick}} \bibnamefont{and}
  \bibinfo{author}{\bibfnamefont{H.}~\bibnamefont{Smith}},
  \eprint{{\it{Bose-Einstein Condensation in Dilute Gases}} (Cambridge
  University Press, Cambridge, UK, 2002)}.

\bibitem[{\citenamefont{Burnett et~al.}(2002)\citenamefont{Burnett, Julienne,
  Lett, Tiesinga, and Williams}}]{BuNa02}
\bibinfo{author}{\bibfnamefont{K.}~\bibnamefont{Burnett}},
  \bibinfo{author}{\bibfnamefont{P.~S.} \bibnamefont{Julienne}},
  \bibinfo{author}{\bibfnamefont{P.~D.} \bibnamefont{Lett}},
  \bibinfo{author}{\bibfnamefont{E.}~\bibnamefont{Tiesinga}}, \bibnamefont{and}
  \bibinfo{author}{\bibfnamefont{C.~J.} \bibnamefont{Williams}},
  \bibinfo{journal}{Nature} \textbf{\bibinfo{volume}{416}},
  \bibinfo{pages}{225} (\bibinfo{year}{2002}).

\bibitem[{\citenamefont{Greiner et~al.}(2002)\citenamefont{Greiner, Mandel,
  Esslinger, H{\"a}nsch, and Bloch}}]{GrNat02}
\bibinfo{author}{\bibfnamefont{M.}~\bibnamefont{Greiner}},
  \bibinfo{author}{\bibfnamefont{O.}~\bibnamefont{Mandel}},
  \bibinfo{author}{\bibfnamefont{T.}~\bibnamefont{Esslinger}},
  \bibinfo{author}{\bibfnamefont{T.~W.} \bibnamefont{H{\"a}nsch}},
  \bibnamefont{and} \bibinfo{author}{\bibfnamefont{I.}~\bibnamefont{Bloch}},
  \bibinfo{journal}{Nature} \textbf{\bibinfo{volume}{415}}, \bibinfo{pages}{39}
  (\bibinfo{year}{2002}).

\bibitem[{\citenamefont{Widera et~al.}(2005)\citenamefont{Widera, Gerbier,
  Foelling, Gericke, Mandel, and Bloch}}]{Bloch1}
\bibinfo{author}{\bibfnamefont{A.}~\bibnamefont{Widera}},
  \bibinfo{author}{\bibfnamefont{F.}~\bibnamefont{Gerbier}},
  \bibinfo{author}{\bibfnamefont{S.}~\bibnamefont{Folling}},
  \bibinfo{author}{\bibfnamefont{T.}~\bibnamefont{Gericke}},
  \bibinfo{author}{\bibfnamefont{O.}~\bibnamefont{Mandel}}, \bibnamefont{and}
  \bibinfo{author}{\bibfnamefont{I.}~\bibnamefont{Bloch}},
  \bibinfo{journal}{Phys. Rev. Lett.} \textbf{\bibinfo{volume}{95}},
  \bibinfo{pages}{190405} (\bibinfo{year}{2005}).

\bibitem[{\citenamefont{Gerbier et~al.}(2006)\citenamefont{Gerbier, Foelling,
  A.~Widera, and Bloch}}]{Bloch2}
\bibinfo{author}{\bibfnamefont{F.}~\bibnamefont{Gerbier}},
  \bibinfo{author}{\bibfnamefont{S.}~\bibnamefont{Folling}},
  \bibinfo{author}{\bibfnamefont{A.}~\bibnamefont{Widera}},
  \bibinfo{author}{\bibfnamefont{O.}~\bibnamefont{Mandel}},
  \bibnamefont{and} \bibinfo{author}{\bibfnamefont{I.}~\bibnamefont{Bloch}},
  \bibinfo{journal}{Phys.\ Rev.\ Lett} \textbf{\bibinfo{volume}{96}},
  \bibinfo{pages}{090401} (\bibinfo{year}{2006}).

\bibitem[{\citenamefont{Demler and Zhou}(2002)}]{DePRL02}
\bibinfo{author}{\bibfnamefont{E.}~\bibnamefont{Demler}} \bibnamefont{and}
  \bibinfo{author}{\bibfnamefont{F.}~\bibnamefont{Zhou}},
  \bibinfo{journal}{Phys. Rev. Lett.} \textbf{\bibinfo{volume}{88}},
  \bibinfo{pages}{163001} (\bibinfo{year}{2002}).

\bibitem[{\citenamefont{A.~Imambekov and Demler}(2003)}]{ImPRA03}
\bibinfo{author}{\bibfnamefont{A.}~\bibnamefont{Imambekov}},
  \bibinfo{author}{\bibfnamefont{M.}~\bibnamefont{Lukin}},
  \bibnamefont{and} \bibinfo{author}{\bibfnamefont{E.}~\bibnamefont{Demler}},
  \bibinfo{journal}{Phys. Rev. A} \textbf{\bibinfo{volume}{68}},
  \bibinfo{pages}{063602} (\bibinfo{year}{2003}).

\bibitem[{\citenamefont{Zhou and Snoek}(2003)}]{ZhAnn03}
\bibinfo{author}{\bibfnamefont{F.}~\bibnamefont{Zhou}} \bibnamefont{and}
  \bibinfo{author}{\bibfnamefont{M.}~\bibnamefont{Snoek}},
  \bibinfo{journal}{Ann. Phys. (NY)} \textbf{\bibinfo{volume}{308}},
  \bibinfo{pages}{692} (\bibinfo{year}{2003}).

\bibitem[{\citenamefont{Snoek and Zhou}(2004)}]{SnPRB04}
\bibinfo{author}{\bibfnamefont{M.}~\bibnamefont{Snoek}} \bibnamefont{and}
  \bibinfo{author}{\bibfnamefont{F.}~\bibnamefont{Zhou}},
  \bibinfo{journal}{Phys. Rev. B} \textbf{\bibinfo{volume}{69}},
  \bibinfo{pages}{094410} (\bibinfo{year}{2004}).

\bibitem[{\citenamefont{Yip}(2003)}]{YPRL03}
\bibinfo{author}{\bibfnamefont{S.~K.} \bibnamefont{Yip}},
  \bibinfo{journal}{Phys. Rev. Lett.} \textbf{\bibinfo{volume}{90}},
  \bibinfo{pages}{250402} (\bibinfo{year}{2003}).

\bibitem[{\citenamefont{Kimura et~al.}()\citenamefont{Kimura, Tsuchiya,
  Yamashita, and Kurihara}}]{Kimura}
\bibinfo{author}{\bibfnamefont{T.}~\bibnamefont{Kimura}},
  \bibinfo{author}{\bibfnamefont{S.}~\bibnamefont{Tsuchiya}},
  \bibinfo{author}{\bibfnamefont{M.}~\bibnamefont{Yamashita}},
  \bibnamefont{and} \bibinfo{author}{\bibfnamefont{S.}~\bibnamefont{Kurihara}},
  \eprint{condmat/0506466}.

\bibitem[{\citenamefont{A.~Imambekov and Demler}(2004)}]{ImPRL04}
\bibinfo{author}{\bibfnamefont{A.}~\bibnamefont{Imambekov}},
  \bibinfo{author}{\bibfnamefont{M.}~\bibnamefont{Lukin}},
  \bibnamefont{and} \bibinfo{author}{\bibfnamefont{E.}~\bibnamefont{Demler}},
  \bibinfo{journal}{Phys. Rev. Lett.} \textbf{\bibinfo{volume}{93}},
  \bibinfo{pages}{120405} (\bibinfo{year}{2004}).

\bibitem[{\citenamefont{Zhou et~al.}(2004)\citenamefont{Zhou, Snoek, Wiemer,
  and Affleck}}]{ZhPRB04}
\bibinfo{author}{\bibfnamefont{F.}~\bibnamefont{Zhou}},
  \bibinfo{author}{\bibfnamefont{M.}~\bibnamefont{Snoek}},
  \bibinfo{author}{\bibfnamefont{J.}~\bibnamefont{Wiemer}}, \bibnamefont{and}
  \bibinfo{author}{\bibfnamefont{I.}~\bibnamefont{Affleck}},
  \bibinfo{journal}{Phys. Rev. B} \textbf{\bibinfo{volume}{70}},
  \bibinfo{pages}{184434} (\bibinfo{year}{2004}).

\bibitem[{\citenamefont{Rizzi et~al.}(2005)\citenamefont{Rizzi, Rossini,
  De~Chiara, Montangero, and Fazio}}]{Rizzi}
\bibinfo{author}{\bibfnamefont{M.}~\bibnamefont{Rizzi}},
  \bibinfo{author}{\bibfnamefont{D.}~\bibnamefont{Rossini}},
  \bibinfo{author}{\bibfnamefont{G.}~\bibnamefont{De~Chiara}},
  \bibinfo{author}{\bibfnamefont{S.}~\bibnamefont{Montangero}},
  \bibnamefont{and} \bibinfo{author}{\bibfnamefont{R.}~\bibnamefont{Fazio}},
  \bibinfo{journal}{Phys. Rev. Lett.} \textbf{\bibinfo{volume}{95}},
  \bibinfo{pages}{240404} (\bibinfo{year}{2005}).

\bibitem[{\citenamefont{K{\"u}hner and Monien}(1998)}]{KuPRB98}
\bibinfo{author}{\bibfnamefont{T.~D.} \bibnamefont{K{\"u}hner}}
  \bibnamefont{and} \bibinfo{author}{\bibfnamefont{H.}~\bibnamefont{Monien}},
  \bibinfo{journal}{Phys. Rev. B.} \textbf{\bibinfo{volume}{58}},
  \bibinfo{pages}{R14741} (\bibinfo{year}{1998}).

\bibitem[{\citenamefont{Jaksch et~al.}(1998)\citenamefont{Jaksch, Bruder,
  Cirac, Gardiner, and Zoller}}]{JaPRL98}
\bibinfo{author}{\bibfnamefont{D.}~\bibnamefont{Jaksch}},
  \bibinfo{author}{\bibfnamefont{C.}~\bibnamefont{Bruder}},
  \bibinfo{author}{\bibfnamefont{J.~I.} \bibnamefont{Cirac}},
  \bibinfo{author}{\bibfnamefont{C.~W.} \bibnamefont{Gardiner}},
  \bibnamefont{and} \bibinfo{author}{\bibfnamefont{P.}~\bibnamefont{Zoller}},
  \bibinfo{journal}{Phys.\ Rev.\ Lett} \textbf{\bibinfo{volume}{81}},
  \bibinfo{pages}{3108} (\bibinfo{year}{1998}).

\bibitem[{\citenamefont{Fisher et~al.}(1989)\citenamefont{Fisher, Weichman,
  Grinstein, and Fisher}}]{FiPRB89}
\bibinfo{author}{\bibfnamefont{M.~P.~A.} \bibnamefont{Fisher}},
  \bibinfo{author}{\bibfnamefont{P.~B.} \bibnamefont{Weichman}},
  \bibinfo{author}{\bibfnamefont{G.}~\bibnamefont{Grinstein}},
  \bibnamefont{and} \bibinfo{author}{\bibfnamefont{D.~S.}
  \bibnamefont{Fisher}}, \bibinfo{journal}{Phys.\ Rev.\ B}
  \textbf{\bibinfo{volume}{40}}, \bibinfo{pages}{546} (\bibinfo{year}{1989}).

\bibitem[{\citenamefont{K{\"u}hner et~al.}(2000)\citenamefont{K{\"u}hner,
  White, and Monien}}]{KuPRB00}
\bibinfo{author}{\bibfnamefont{T.~D.} \bibnamefont{K{\"u}hner}},
  \bibinfo{author}{\bibfnamefont{S.~R.} \bibnamefont{White}}, \bibnamefont{and}
  \bibinfo{author}{\bibfnamefont{H.}~\bibnamefont{Monien}},
  \bibinfo{journal}{Phys. Rev. B.} \textbf{\bibinfo{volume}{61}},
  \bibinfo{pages}{12474} (\bibinfo{year}{2000}).

\bibitem[{\citenamefont{Haldane}(1983)}]{HaPRL83}
\bibinfo{author}{\bibfnamefont{F.}~\bibnamefont{Haldane}},
  \bibinfo{journal}{Phys Lett.} \textbf{\bibinfo{volume}{93A}},
  \bibinfo{pages}{464} (\bibinfo{year}{1983}).

\bibitem{DimerOrder}
\bibinfo{author}{\bibfnamefont{J.}~\bibnamefont{Parkinson}},
  \bibinfo{journal}{J. Phys. C} \textbf{\bibinfo{volume}{20}},
  \bibinfo{pages}{L1029} (\bibinfo{year}{1988});
%\bibitem[{\citenamefont{Barber and Batchelor}(1989)}]{BaPRB89}
\bibinfo{author}{\bibfnamefont{M.~N.} \bibnamefont{Barber}} \bibnamefont{and}
  \bibinfo{author}{\bibfnamefont{M.~T.} \bibnamefont{Batchelor}},
  \bibinfo{journal}{Phys. Rev. B} \textbf{\bibinfo{volume}{40}},
  \bibinfo{pages}{4621} (\bibinfo{year}{1989});
%\bibitem[{\citenamefont{Kl{\"u}mper}(1990)}]{KlEPL90}
\bibinfo{author}{\bibfnamefont{A.}~\bibnamefont{Kl{\"u}mper}},
  \bibinfo{journal}{Eruophys. Lett.} \textbf{\bibinfo{volume}{9}},
  \bibinfo{pages}{815} (\bibinfo{year}{1990});
%\bibitem[{\citenamefont{Xian}(1993)}]{XiPLA93}
\bibinfo{author}{\bibfnamefont{Y.}~\bibnamefont{Xian}}, \bibinfo{journal}{Phys.
  Lett. A} \textbf{\bibinfo{volume}{183}}, \bibinfo{pages}{437}
  (\bibinfo{year}{1993}).

\bibitem[{\citenamefont{Chubukov}(1991)}]{ChubDimer}
\bibinfo{author}{\bibfnamefont{A.~V.}~\bibnamefont{Chubukov}},
  \bibinfo{journal}{Phys. Rev. B} \textbf{\bibinfo{volume}{43}},
  \bibinfo{pages}{3337} (\bibinfo{year}{1991});
%\bibitem[{\citenamefont{Chubukov}(1990)}]{ChubJPC90}
%\bibinfo{author}{\bibfnamefont{A.~V.}~\bibnamefont{Chubukov}},
  \bibinfo{journal}{J. Phys.; Condens. Matter} \textbf{\bibinfo{volume}{2}},
  \bibinfo{pages}{1593} (\bibinfo{year}{1990}).

\bibitem[{\citenamefont{Buchta et~al.}(2005)\citenamefont{Buchta, F{\'a}th,
  Legeza, and S{\'o}lyom}}]{BuPRB05}
\bibinfo{author}{\bibfnamefont{K.}~\bibnamefont{Buchta}},
  \bibinfo{author}{\bibfnamefont{G.}~\bibnamefont{F{\'a}th}},
  \bibinfo{author}{\bibfnamefont{{\"O}.}~\bibnamefont{Legeza}},
  \bibnamefont{and}
  \bibinfo{author}{\bibfnamefont{J.}~\bibnamefont{S{\'o}lyom}},
  \bibinfo{journal}{Phys. Rev. B} \textbf{\bibinfo{volume}{72}},
  \bibinfo{pages}{054433} (\bibinfo{year}{2005}).

\bibitem[{\citenamefont{L{\"a}uchli et~al.}()\citenamefont{L{\"a}uchli, Schmid,
  and Trebst}}]{Lauchli}
\bibinfo{author}{\bibfnamefont{A.}~\bibnamefont{L{\"a}uchli}},
  \bibinfo{author}{\bibfnamefont{G.}~\bibnamefont{Schmid}}, \bibnamefont{and}
  \bibinfo{author}{\bibfnamefont{S.}~\bibnamefont{Trebst}},
  \eprint{condmat/0607173}.

\bibitem[{\citenamefont{Porras et~al.}(2006)\citenamefont{Porras, Verstraete,
  and Cirac}}]{Verstraete}
\bibinfo{author}{\bibfnamefont{D.}~\bibnamefont{Porras}},
  \bibinfo{author}{\bibfnamefont{F.}~\bibnamefont{Verstraete}},
  \bibnamefont{and} \bibinfo{author}{\bibfnamefont{J.~I.}~\bibnamefont{Cirac}},
  \bibinfo{journal}{Phys. Rev. B} \textbf{\bibinfo{volume}{73}},
  \bibinfo{pages}{014410} (\bibinfo{year}{2006}).

\bibitem[{\citenamefont{Tsuchiya et~al.}(2004)\citenamefont{Tsuchiya, Kimura,
  Yamashita, and Kurihara}}]{TsPRA04}
\bibinfo{author}{\bibfnamefont{S.}~\bibnamefont{Tsuchiya}},
  \bibinfo{author}{\bibfnamefont{S.}~\bibnamefont{Kurihara}},
  \bibnamefont{and} \bibinfo{author}{\bibfnamefont{T.}~\bibnamefont{Kimura}},
  \bibinfo{journal}{Phys. Rev. A} \textbf{\bibinfo{volume}{70}},
  \bibinfo{pages}{043628} (\bibinfo{year}{2004}).

\bibitem{WhiteDMRG}
\bibinfo{author}{\bibfnamefont{S.~R.} \bibnamefont{White}},
  \bibinfo{journal}{Phys. Rev. Lett.} \textbf{\bibinfo{volume}{69}},
  \bibinfo{pages}{2863} (\bibinfo{year}{1992});
%\bibitem[{\citenamefont{White}(1993)}]{WhitePRB92}
\bibinfo{author}{\bibfnamefont{S.~R.} \bibnamefont{White}},
  \bibinfo{journal}{Phys. Rev. B} \textbf{\bibinfo{volume}{48}},
  \bibinfo{pages}{10345} (\bibinfo{year}{1993}).

\bibitem[{\citenamefont{Schollw\"ock}(2005)}]{UliReview05}
\bibinfo{author}{\bibfnamefont{U.}~\bibnamefont{Schollw\"ock}},
  \bibinfo{journal}{Rev. Mod. Phys.} \textbf{\bibinfo{volume}{77}},
  \bibinfo{eid}{259} (\bibinfo{year}{2005}).

\bibitem[{\citenamefont{McCulloch and Gul\'acsi}(2002)}]{NonAbelianDMRG02}
\bibinfo{author}{\bibfnamefont{I.~P.} \bibnamefont{McCulloch}}
  \bibnamefont{and}
  \bibinfo{author}{\bibfnamefont{M.}~\bibnamefont{Gul\'acsi}},
  \bibinfo{journal}{Europhys. Lett.} \textbf{\bibinfo{volume}{57}},
  \bibinfo{pages}{852} (\bibinfo{year}{2002}).

\bibitem[{\citenamefont{Biedenharn and Louck}(1981)}]{SU2Intro}
See, for example:
\bibinfo{author}{\bibfnamefont{L.~C.} \bibnamefont{Biedenharn}}
  \bibnamefont{and} \bibinfo{author}{\bibfnamefont{J.~D.} \bibnamefont{Louck}},
  \emph{\bibinfo{title}{Angular Momentum in Quantum Physics}}
  (\bibinfo{publisher}{Addison Wesley, Massachusetts}, \bibinfo{year}{1981}).

\bibitem[{\citenamefont{Wilson}(1975)}]{WilsonNRG}
\bibinfo{author}{\bibfnamefont{K.~G.} \bibnamefont{Wilson}},
  \bibinfo{journal}{Rev. Mod. Phys.} \textbf{\bibinfo{volume}{47}},
  \bibinfo{pages}{773} (\bibinfo{year}{1975}).

\bibitem[{\citenamefont{Haldane}(1981)}]{HaPRL81}
\bibinfo{author}{\bibfnamefont{F.~D.~M.}~\bibnamefont{Haldane}},
  \bibinfo{journal}{Phys. Rev. Lett} \textbf{\bibinfo{volume}{47}},
  \bibinfo{pages}{1840} (\bibinfo{year}{1981}).

\bibitem[{\citenamefont{Zhou}(2003)}]{ZhEPL03}
\bibinfo{author}{\bibfnamefont{F.}~\bibnamefont{Zhou}},
  \bibinfo{journal}{Europhys. Lett.} \textbf{\bibinfo{volume}{63}},
  \bibinfo{pages}{505} (\bibinfo{year}{2003}).

\bibitem[{\citenamefont{Zhou}(2003)}]{SchmidtParabolic}
\bibinfo{author}{\bibfnamefont{H.~-J.} \bibnamefont{Schmidt}},
  \bibinfo{author}{\bibfnamefont{J.} \bibnamefont{Schnack}}
  \bibnamefont{and}
  \bibinfo{author}{\bibfnamefont{M.}~\bibnamefont{Luban}},
  \bibinfo{journal}{Europhys. Lett.} \textbf{\bibinfo{volume}{55}},
  \bibinfo{pages}{105} (\bibinfo{year}{2001}).

\end{thebibliography}
\end{document}